\documentclass[twocolumn,english,prl,showpacs,superscriptaddress]{revtex4}

\usepackage[T1]{fontenc}
\usepackage{color}
\usepackage{babel}
\usepackage{amsmath}
\usepackage{amssymb}
\usepackage{graphicx}
\usepackage{esint}
\usepackage[unicode=true,pdfusetitle,
 bookmarks=true,bookmarksnumbered=false,bookmarksopen=false,
 breaklinks=false,pdfborder={0 0 1},backref=false,colorlinks=true]
 {hyperref}
\hypersetup{
 citecolor=blue}
\usepackage{breakurl}

\makeatletter
\@ifundefined{textcolor}{}
{%
 \definecolor{BLACK}{gray}{0}
 \definecolor{WHITE}{gray}{1}
 \definecolor{RED}{rgb}{1,0,0}
 \definecolor{GREEN}{rgb}{0,1,0}
 \definecolor{BLUE}{rgb}{0,0,1}
 \definecolor{CYAN}{cmyk}{1,0,0,0}
 \definecolor{MAGENTA}{cmyk}{0,1,0,0}
 \definecolor{YELLOW}{cmyk}{0,0,1,0}
}

\makeatother

\begin{document}

\title{Quantum optics in a non-inertial reference frame: the Rabi splitting
in a rotating ring cavity}

\author{Sheng-Wen Li}

\affiliation{Beijing Computational Science Research Center, Beijing 100084, China}

\affiliation{Synergetic Innovation Center of Quantum Information and Quantum Physics,
University of Science and Technology of China, Hefei 230026, China}

\author{Z. H. Wang}

\affiliation{Beijing Computational Science Research Center, Beijing 100084, China}

\affiliation{Center for Quantum Sciences, Northeast Normal University, Changchun
130117, China}

\author{Lan Zhou}

\affiliation{Synergetic Innovation Center of Quantum\emph{ }Effects and Applications,
Department of Physics, Hunan Normal University, Changsha 410081, China}

\author{C. P. Sun}

\email{cpsun@csrc.ac.cn}

\homepage{http://www.csrc.ac.cn/~suncp}

\selectlanguage{english}%

\affiliation{Beijing Computational Science Research Center, Beijing 100084, China}

\affiliation{Synergetic Innovation Center of Quantum Information and Quantum Physics,
University of Science and Technology of China, Hefei 230026, China}
\begin{abstract}
We study quantum optics with the atoms coupled to the quantized electromagnetic
(EM) field in a non-inertial reference frame by making use of quantum
field theory in curved spacetime. We rigorously establish the microscopic
model for a two-level atom interacting with the quantized EM field
in a rotating ring cavity by deriving a Jaynes-Cummings (JC) type
Hamiltonian. Due to the two fold degeneracy of the ring cavity modes,
the Rabi splitting exhibits three rather than two resonant frequency
peaks. We find that the heights of the two side peaks show a sensitive
linear dependence on the rotating velocity. This high sensitivity
can be utilized to detect the angular velocity of the whole system.
\end{abstract}


\maketitle
\textbf{Introduction}\emph{ --} The interference of of two light beams
in a rotating ring can be utilized to measure the rotating velocity.
This is well known as the Sagnac effect, which bases some optical
gyroscope schemes \cite{post_sagnac_1967,chow_ring_1985,zimmer_sagnac_2004,steinberg_rotating_2005,shahriar_ultrahigh_2007,gu_sagnac_2011}.
For the Sagnac effect in the medium with linear dispersion, there
have been a lot of sophisticated studies based on classical optics.
If we want to make the optical gyroscopes to an extremely high precision,
it is necessary to consider the quantum fluctuations in this rotating
optical system. We notice that a rigorous quantum theory about the
microscopic model about the interaction between the atoms and the
quantized EM field in a rotating reference frame is still not well
established, but it is obviously essential for the study of quantum
and  nonlinear effects in a rotating optical system.

In this letter, we ascribe the effects of rotation to the ``curved''
spacetime metric according to the generic principles in relativity.
Starting from the classical Lagrangians of the EM field and a charged
particle based on the the principle of least action, we obtain the
covariant motion equations in the rotating non-inertial reference
frame (NIRF). The variation is carried out in the rotating reference
frame, and all the non-inertial physical effects rooted in the rotation
are included in the ``curved'' spacetime metric. Then we obtain
the quantized Hamiltonian for quantum optics in NIRF through the canonical
quantization.

We derive a microscopic model of an atom interacting with the quantized
EM field in a rotating reference frame, which then gives the Jaynes-Cummings
(JC) model for a rotating ring cavity coupled with a two-level atom.
For a ring cavity in the inertial rest frame, the clockwise (CW) and
counter-clockwise (CCW) propagating optical modes are always exactly
degenerated. Thus, in the JC model of a ring cavity, the atom couples
with the two degenerate modes simultaneously. Due to the existence
of two optical modes, the Rabi splitting of this system exhibits three
rather than two resonant frequency peaks. More importantly, we find
that the rotation of the ring would induce a detuning between the
original degenerated modes in the non-inertial frame, and the heights
of the two side peaks would change with the rotating speed due to
this rotation induced detuning. At lower speed, the heights of the
side peaks depend linearly on the rotating velocity. This sensitivity
can be utilized to detect the rotation of the whole system, which
can be regarded as a quantum Sagnac effect.

\begin{figure}
\includegraphics[width=0.65\columnwidth]{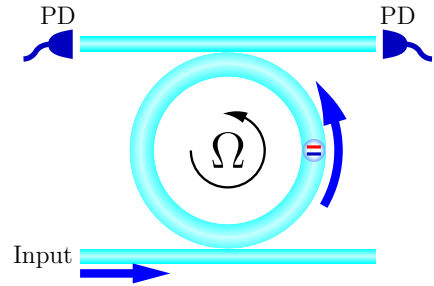}

\protect\caption{(Color online) Schematic setup. A two-level atom is fixed in the rotating
ring, and external fibers are used for driving and probing. Two photon
detectors (PD) are used to measure the photon current.}

\label{fig-setup}
\end{figure}

\textbf{EM field in the rotating reference frame}\emph{ -- }We consider
the EM field rotating around the $z$-axial in the CCW direction with
angular speed $\Omega$. Physics laws must have the same covariant
mathematical form in all reference frames (including the non-inertial
rotating frame that we are studying). Thus, in the rotating reference
frame, the Lagrangian density of the EM field is 
\begin{equation}
{\cal L}=-\frac{1}{4\mu_{0}}F^{\mu\nu}F_{\mu\nu}.
\end{equation}
Here $\mu_{0}$ is the magnetic constant, $F_{\mu\nu}:=\nabla_{\mu}A_{\nu}-\nabla_{\nu}A_{\mu}=\partial_{\mu}A_{\nu}-\partial_{\nu}A_{\mu}$
and $F^{\mu\nu}:=g^{\mu\alpha}g^{\nu\beta}F_{\alpha\beta}$. In the
above definitions, $\partial_{\mu}:=\partial/\partial x^{\mu}$, where
 $x^{\mu}:=(ct,x,y,z)$ is the 4-dimensional coordinate, and $A_{\mu}:=(-\phi/c,\mathbf{A}_{i})$
($i=1,2,3$) is the electromagnetic 4-potential. $\nabla_{\mu}$ is
the covariant derivation defined by
\begin{equation}
\nabla_{\mu}A_{\nu}=\partial_{\mu}A_{\nu}-\Gamma_{\mu\nu}^{\tau}A_{\tau}.
\end{equation}
Here $\Gamma_{\mu\nu}^{\tau}$ is the Christoffel symbol and it can
be calculated from the spacetime metric $g^{\mu\nu}$ \cite{weinberg_gravitation_1972,misner_gravitation_1973}.
All the physical effects due to the rotation are included in the ``curved''
spacetime metric $g^{\mu\nu}$ (see Ref.\,\cite{menegozzi_theory_1973,chow_ring_1985}
or Appendix \ref{sub:Space-time-metric}), i.e., 
\begin{equation}
g^{\mu\nu}=\left[\begin{array}{cccc}
-1 & -\frac{\Omega y}{c} & \frac{\Omega x}{c}\\
-\frac{\Omega y}{c} & 1-\frac{\Omega^{2}y^{2}}{c^{2}} & \frac{\Omega^{2}xy}{c^{2}}\\
\frac{\Omega x}{c} & \frac{\Omega^{2}xy}{c^{2}} & 1-\frac{\Omega^{2}x^{2}}{c^{2}}\\
 &  &  & 1
\end{array}\right].\label{eq:g^uv}
\end{equation}

From the variation of the action 
\begin{equation}
\delta S=\delta\int{\cal L}\cdot\sqrt{-\mathfrak{g}}d^{4}x^{\mu}:=\delta\int\tilde{{\cal L}}\,d^{4}x^{\mu},
\end{equation}
 we obtain the Euler-Lagrangian equation as
\begin{gather*}
\partial_{\mu}\Big[\frac{\partial\tilde{{\cal L}}}{\partial(\partial_{\mu}A_{\nu})}\Big]=\frac{\partial\tilde{{\cal L}}}{\partial A_{\nu}}\Rightarrow\frac{1}{\sqrt{-\mathfrak{g}}}\partial_{\mu}[\sqrt{-\mathfrak{g}}F^{\mu\nu}]=\nabla_{\mu}F^{\mu\nu}=0,
\end{gather*}
where $\mathfrak{g}:=\det g^{\mu\nu}$. Here $\tilde{{\cal L}}:=\sqrt{-\mathfrak{g}}{\cal L}$
is a functional of $A_{\mu}$ and $\partial_{\mu}A_{\nu}$. This is
the covariant form of the Maxwell equation, which applies to all general
reference frames (both inertial and non-inertial ones) \cite{weinberg_gravitation_1972}.
Substituting the spacetime metric Eq.\,(\ref{eq:g^uv}) into the
above covariant equation, and taking the Coulomb gauge $\nabla\cdot\mathbf{A}=0,\,A_{0}=0$,
we obtain the d'Alembert equation $\nabla^{\mu}\nabla_{\mu}\mathbf{A}_{i}=0$
as 
\begin{equation}
[-\partial_{0}^{2}+\nabla^{2}+2\tilde{\mathbf{v}}_{R}\cdot\nabla\partial_{0}-(\tilde{\mathbf{v}}_{R}\cdot\nabla)^{2}]\mathbf{A}_{i}=0.
\end{equation}
Here $\partial_{0}:=c^{-1}\partial/\partial t$, $\tilde{\mathbf{v}}_{R}:=\mathbf{v}_{R}/c$,
and $\mathbf{v}_{R}:=\boldsymbol{\Omega}\times\mathbf{r}$ is the
linear velocity.

Next we reduce the problem into a quasi-1D ring configuration. We
assume that $\mathbf{A}(\mathbf{r})$ is homogenous in the transverse
direction, and we have $\nabla\simeq\hat{\mathrm{e}}_{s}\partial_{s}$
\cite{menegozzi_theory_1973}, where $\hat{\mathrm{e}}_{s}$ is the
direction along the ring. Then we have
\begin{equation}
[-\partial_{0}^{2}+2\tilde{v}_{R}\partial_{s}\partial_{0}+(1-\tilde{v}_{R}^{2})\partial_{s}^{2}]\mathbf{A}_{i}=0.\label{eq:motion}
\end{equation}
Here $\tilde{v}_{R}=v_{R}/c$ and $v_{R}:=|\mathbf{v}_{R}|=\Omega R$
is the linear speed. The general solution of the above equation is
\begin{equation}
\mathbf{A}(s,t)=\sum_{k,\lambda}\hat{\mathrm{e}}_{k\lambda}\overline{\mathrm{Z}}_{k}[\alpha_{k\lambda}e^{iks-i\omega_{k}t}+\alpha_{k\lambda}^{*}e^{-iks+i\omega_{k}t}],
\end{equation}
where $k=2\pi n/\mathsf{L},\,n\in\mathbb{Z}$, and $\mathsf{L}$ is
the length of the ring  \footnote{Strictly speaking, when the ring is rotating, the length is $\mathsf{L}=2\pi R\cdot[1-(v_{R}/c)^{2}]^{-1}>2\pi R$.
Here we omit this small change when $v_{R}/c\ll1$, because it is
 $\sim o(v_{R}^{2}/c^{2})$.}; $\hat{\mathrm{e}}_{k\lambda}$ is the polarization directions in
the transverse section, and $\overline{\mathrm{Z}}_{k}$ is a normalization
constant. The eigenvalue equation of Eq.\,(\ref{eq:motion}), $\omega_{k}^{2}+2v_{R}k\omega_{k}-(c^{2}-v_{R}^{2})k^{2}=0$,
gives rise to the following dispersion relation 
\begin{equation}
\omega_{k}=\pm ck-v_{R}k.\label{eq:disperse}
\end{equation}
Therefore, the rotation makes the dispersion relation anisotropic,
i.e., for the two modes $k_{+}=|k|$ and $k_{-}=-|k|$, their frequencies
$\omega_{k_{\pm}}$ no longer equal.

\textbf{Quantization of the EM field}\textbf{\emph{ }}-- Under the
Coulomb gauge we used above, the canonical momentums are ${\cal E}^{0}=0$
and 
\begin{equation}
{\cal E}^{i}=\frac{\partial\tilde{{\cal L}}}{\partial(\partial_{t}A_{i})}=-\varepsilon_{0}(\mathbf{E}-\mathbf{v}_{R}\times\mathbf{B})_{i},
\end{equation}
 where $\varepsilon_{0}$ is the electric constant. The Hamiltonian
density ${\cal H}={\cal E}^{i}\partial_{t}A_{i}-\tilde{{\cal L}}$
is obtained as 
\begin{equation}
{\cal H}=\frac{1}{2}(\varepsilon_{0}\mathbf{E}^{2}+\frac{1}{\mu_{0}}\mathbf{B}^{2})-\frac{1}{2}\varepsilon_{0}(\mathbf{v}_{R}\times\mathbf{B})^{2}.
\end{equation}
 Restricted in the quasi-1D ring configuration, respectively they
become
\begin{align}
{\cal E}^{i} & =\varepsilon_{0}(\partial_{t}\mathbf{A}-v_{R}\partial_{s}\mathbf{A})_{i},\nonumber \\
{\cal H} & =\frac{1}{2}\varepsilon_{0}(\partial_{t}\mathbf{A})^{2}+\frac{1-\tilde{v}_{R}^{2}}{2\mu_{0}}(\partial_{s}\mathbf{A})^{2}.
\end{align}
Notice that since $\nabla\simeq\hat{\mathrm{e}}_{s}\partial_{s}$,
the Coulomb gauge $\nabla\cdot\mathbf{A}=0$ leads to $\mathbf{A}\cdot\hat{\mathrm{e}}_{s}=0$.
Namely, $\mathbf{A}(s,t)$ only has two transverse directions, and
so does the canonical momentum $\boldsymbol{{\cal E}}(s,t)$.

We apply the following canonical quantization condition
\begin{align}
[\hat{\mathbf{A}}_{\lambda}(s,t),\,\hat{{\cal E}}^{\sigma}(s',t)] & =i\hbar\delta_{\lambda}^{\sigma}\delta(s-s')\cdot\mathsf{S}^{-1},\label{eq:quant}\\
{}[\hat{\mathbf{A}}_{\lambda}(s,t),\,\hat{\mathbf{A}}_{\sigma}(s',t)] & =[\hat{{\cal E}}^{\lambda}(s,t),\,\hat{{\cal E}}^{\sigma}(s',t)]=0.\nonumber 
\end{align}
Here $\lambda,\sigma=1,2$ mean the two transverse directions, and
$\mathsf{S}$ is the cross-sectional area of the ring. Since we have
reduced the problem into 1-dimension the above quantization condition
is consistent with the Coulomb gauge condition $\nabla\cdot\mathbf{A}=0$
automatically. Now we write down the field operator $\hat{\mathbf{A}}(s,t)$
as 
\begin{equation}
\hat{\mathbf{A}}(s,t)=\sum_{k,\lambda}\hat{\mathrm{e}}_{k\lambda}\overline{\mathrm{Z}}_{k}[\hat{a}_{k\lambda}e^{iks-i\omega_{k}t}+\hat{a}_{k\lambda}^{\dagger}e^{-iks+i\omega_{k}t}].\label{eq:A(s,t)}
\end{equation}
With the help of the above canonical quantization condition (\ref{eq:quant}),
we can prove the following bosonic commutation relations (see Appendix
\ref{sec:Canonical-quantization-of}),
\begin{align}
[\hat{a}_{k\lambda},\,\hat{a}_{q\sigma}] & =[\hat{a}_{k\lambda}^{\dagger},\,\hat{a}_{q\sigma}^{\dagger}]=0,\nonumber \\
{}[\hat{a}_{k\lambda},\,\hat{a}_{q\sigma}^{\dagger}] & =\delta_{kq}\delta_{\lambda\sigma}\cdot\frac{\hbar\cdot|\overline{\mathrm{Z}}_{k}|^{-2}}{2\epsilon_{0}Vc|k|}.
\end{align}
 Thus, the normalization constant is taken as $\overline{\mathrm{Z}}_{k}=\sqrt{\hbar/2\epsilon_{0}Vc|k|}$,
where $V=\mathsf{L}\cdot\mathsf{S}$ is the effective volume of the
ring, so that $[\hat{a}_{k\lambda},\,\hat{a}_{q\sigma}^{\dagger}]=\delta_{kq}\delta_{\lambda\sigma}$.
Then we obtain the quantized Hamiltonian of the EM field as 
\begin{equation}
\hat{H}_{\mathrm{EM}}=\int dV\,{\cal H}=\sum_{k,\lambda}\hbar\omega_{k}(\hat{a}_{k\lambda}^{\dagger}\hat{a}_{k\lambda}+\frac{1}{2}).\label{eq:H-em}
\end{equation}
This Hamiltonian has the same form as that of the EM field in the
inertial frame, but the dispersion relation $\omega_{k}$ is changed
due to the rotation {[}see Eq.\,(\ref{eq:disperse}){]}.

\textbf{JC-model in the rotating frame} -- Next we study the motion
of a charged particle in the EM field in the rotating reference frame.
To this end, we start with the invariant Lagrangian of a charged particle
in the effective curved spacetime \cite{weinberg_gravitation_1972,misner_gravitation_1973}
\begin{equation}
L=-mc[-g_{\mu\nu}\frac{dx^{\mu}}{d\tau}\frac{dx^{\nu}}{d\tau}]^{\frac{1}{2}}+eA_{\mu}\frac{dx^{\mu}}{d\tau},
\end{equation}
where $\tau$ is the proper time. The Hamiltonian description is obtained
from the following variation (see Ref.\,\cite{misner_gravitation_1973}
or Appendix \ref{sec:Hamiltonian-description}) 
\begin{equation}
\delta S=\delta\int\tilde{L}dt:=\delta\int[P_{i}v^{i}-H_{\mathrm{e}}]dt.
\end{equation}
Here $v^{\mu}:=dx^{\mu}/dt$, $v_{\mu}:=dx_{\mu}/dt=g_{\mu\nu}v^{\nu}$,
$\tilde{L}:=L\cdot\Gamma^{-1}$ and
\begin{equation}
\Gamma:=\frac{dt}{d\tau}=\frac{c}{\sqrt{-v_{0}v^{0}-v_{i}v^{i}}}.
\end{equation}
As a conservative system, the Hamiltonian $H_{\mathrm{e}}$ is the
functional of $x^{i}(t)$ and $P_{i}(t)$ ($i=1,2,3$), and 
\begin{align}
P_{i} & =\frac{\partial\tilde{L}}{\partial v^{i}}=\Gamma mv_{i}+eA_{i}:=p_{i}+eA_{i},\nonumber \\
H_{\mathrm{e}} & =P_{i}v^{i}-\tilde{L}=-\Gamma mv_{0}v^{0}-eA_{0}v^{0}.
\end{align}
Here $p_{i}:=\Gamma mv_{i}$ is the mechanical momentum. Replacing
$v_{i}$ and $v^{i}$ by the canonical momentum $P_{i}=p_{i}+eA_{i}$
in $H_{e}$, the Hamiltonian of the charged particle is obtained as
(see Appendix \ref{sec:Hamiltonian-description})
\[
H_{\mathrm{e}}=\frac{v^{0}}{g^{00}}[g^{0i}p_{i}-\sqrt{(g^{0i}p_{i})^{2}-g^{00}(g^{ij}p_{i}p_{j}+m^{2}c^{2})}]-ev^{0}A_{0}.
\]
Substituting in the metric $g^{\mu\nu}$ {[}Eq.\,(\ref{eq:g^uv}){]},
the above Hamiltonian becomes
\begin{align}
H_{\mathrm{e}} & =\sqrt{(\mathbf{P}-e\mathbf{A})^{2}c^{2}+m^{2}c^{4}}+\mathbf{v}_{R}\cdot(\mathbf{P}-e\mathbf{A})+e\varphi\nonumber \\
 & \simeq\frac{1}{2m}(\mathbf{P}-e\mathbf{A})^{2}+\mathbf{v}_{R}\cdot(\mathbf{P}-e\mathbf{A})+e\varphi,
\end{align}
where $\mathbf{P}:=(P_{1},P_{2},P_{3})$. In the quasi-1D ring, $\mathbf{A}$
is always in the transverse direction and perpendicular to the linear
velocity $\mathbf{v}_{R}$, thus $\mathbf{v}_{R}\cdot\mathbf{A}=0$.

We consider an atom with the nucleus fixed at a certain position inside
the ring cavity. Around the nucleus, the electron is trapped by the
central force potential $\varphi\sim r^{-1}$. Thus, in the expansion
of the above equation, $H_{\mathrm{atom}}:=\mathbf{P}^{2}/2m+e\varphi+\mathbf{v}_{R}\cdot\mathbf{P}$
is exactly the Hamiltonian of a hydrogen-like atom plus a perturbation
term $\mathbf{v}_{R}\cdot\mathbf{P}$, which comes from the non-inertial
effect of rotation, and the term $\mathbf{P}\cdot\mathbf{A}$ describes
the coupling between the atom and the EM field.

Treating $\hat{\mathbf{P}}$ and $\hat{\mathbf{x}}$ as operators,
we obtain the energy levels of the hydrogen-like atom. Then we focus
on two energy levels $|\mathsf{e}\rangle$ and $|\mathsf{g}\rangle$
with a dipole transition. Neglecting the $\hat{\mathbf{A}}^{2}$ term,
with the help of Eq.\,(\ref{eq:A(s,t)}), the total Hamiltonian $\hat{H}=\hat{H}_{\mathrm{e}}+\hat{H}_{\mathrm{EM}}$
reads, 
\[
\hat{H}=\frac{\hbar\Omega}{2}\hat{\sigma}^{z}+\xi\hat{\sigma}^{y}+\hat{\sigma}^{y}[\sum_{k,\lambda}\eta_{k\lambda}\hat{a}_{k\lambda}+\eta_{k\lambda}^{*}\hat{a}_{k\lambda}^{\dagger}]+\sum_{k,\lambda}\hbar\omega_{k}\hat{a}_{k\lambda}^{\dagger}\hat{a}_{k\lambda},
\]
 where $\hat{\sigma}^{z}:=|\mathsf{e}\rangle\langle\mathsf{e}|-|\mathsf{g}\rangle\langle\mathsf{g}|$,
$\hat{\sigma}^{x}:=|\mathsf{e}\rangle\langle\mathsf{g}|+|\mathsf{g}\rangle\langle\mathsf{e}|$
and $\hat{\sigma}^{y}=i\hat{[\sigma},^{z}\hat{\sigma}^{x}]$. The
coefficients $\xi$ and $\eta_{k\lambda}$ are explicitly given as
\begin{align}
\xi & =i\langle\mathsf{e}|\mathbf{v}_{R}\cdot\hat{\mathbf{P}}|\mathsf{g}\rangle=-\frac{mv_{R}\Omega}{e}\cdot(\vec{\mathfrak{p}}\cdot\hat{\mathrm{e}}_{s}),\nonumber \\
\eta_{k\lambda} & =\frac{-ie}{m}\langle\mathsf{e}|\overline{\mathrm{Z}}_{k}e^{iks}(\hat{\mathbf{P}}\cdot\hat{\mathrm{e}}_{k\lambda})|\mathsf{g}\rangle\label{eq:coupling}\\
 & \simeq\Omega(\vec{\mathfrak{p}}\cdot\hat{\mathrm{e}}_{k\lambda})[\frac{\hbar}{2\varepsilon_{0}Vc|k|}]^{\frac{1}{2}}e^{iks_{0}},\nonumber 
\end{align}
 where $\vec{\mathfrak{p}}:=e\langle\mathsf{e}|\hat{\mathbf{x}}|\mathsf{g}\rangle$
is the dipole moment \cite{scully_quantum_1997}. Here the dipole
approximation is applied, and $s_{0}$ is the position of the atom.
We choose $s_{0}=0$ to cancel the phase factors. Comparing with the
inertial case $v_{R}=0$, we see that $\Omega$ and $\eta_{k\lambda}$
is unchanged, and the contribution of rotation appears in the correction
term $\xi\hat{\sigma}^{y}$ and the dispersion relation $\omega_{k}$.

We set the two polarization directions to be parallel and vertical
to the projection of $\vec{\mathfrak{p}}$ in the transversal section
respectively. Then the vertical optical modes are decoupled with the
atom {[}see Eq.\,(\ref{eq:coupling}){]}. When the ring is not too
long, the frequencies of different eigen modes of the EM field are
well separated from each other, so we only consider the modes nearly
resonant with the atom. But we should notice that, different from
the Fabry-P\'erot type, the $k_{+}=|k|$ and $k_{-}=-|k|$ modes
are always nearly degenerate in ring cavity (unless $k=0$). When
$v_{R}=0$, they are exactly degenerated $\omega_{k_{+}}=\omega_{k_{-}}$.
That means, the $k_{+}$ and $k_{-}$ modes must be considered together.
Therefore, by omitting the double creation and annihilation terms,
we establish the JC-model of two-level atom in a rotating ring cavity
with the JC Hamiltonian
\begin{align}
\hat{H}= & \frac{\hbar\Omega}{2}\hat{\sigma}^{z}+\xi\hat{\sigma}^{y}+\hbar\omega_{+}\hat{a}_{+}^{\dagger}\hat{a}_{+}+\hbar\omega_{-}\hat{a}_{-}^{\dagger}\hat{a}_{-}\nonumber \\
 & \qquad+g\hat{\sigma}^{+}(\hat{a}_{+}+\hat{a}_{-})+g^{*}\hat{\sigma}^{-}(\hat{a}_{+}^{\dagger}+\hat{a}_{-}^{\dagger}),\label{eq:H_JC}
\end{align}
where the coupling strength $g$ is
\begin{equation}
g=i\Omega(\vec{\mathfrak{p}}\cdot\hat{\mathrm{e}}_{k\lambda})[\frac{\hbar}{2\varepsilon_{0}Vc|k|}]^{\frac{1}{2}}.
\end{equation}
Here $\hat{a}_{\pm}$ is the annihilation operator for the modes $k_{\pm}$,
and their frequencies are $\omega_{\pm}=(c\mp v_{R})|k|:=\omega_{0}\pm\Delta$,
where we define $\omega_{0}:=c|k|$ as the rest frequency of the ring
cavity and $\Delta:=-v_{R}k$ as the rotation detuning. For simplicity,
hereafter we absorb the phase of $g$ into the operators to make $g=g^{*}>0$
and set $\hbar=1$.

\begin{figure}
\includegraphics[width=1\columnwidth]{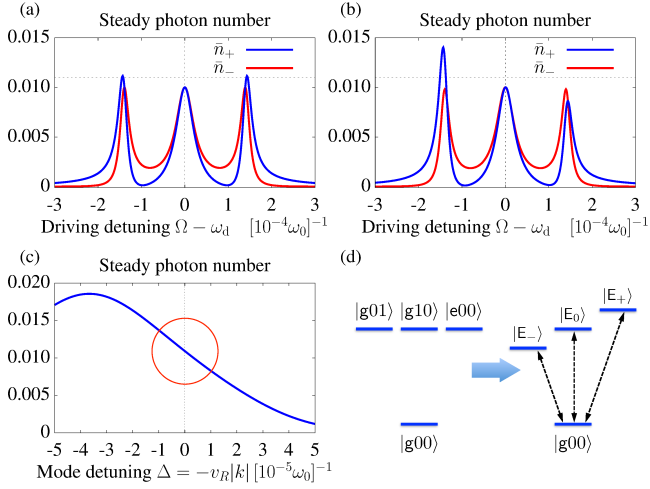}

\protect\caption{(Color online) (a, b) The steady photon number $\overline{n}_{\pm}$
of the two cavity modes. The two-level atom is resonant with the rest
frequency $\Omega=\omega_{0}$, and we set $\omega_{0}=\Omega=1$
as the unit. The other parameters are $g=1\times10^{-4}$, $\gamma=0.5\times10^{-4}$,
${\cal E}=0.05\times10^{-4}$, and (a) $\Delta=0$ (b) $\Delta=-v_{R}k=1\times10^{-5}$.
(c) The height of the side peak of $\overline{n}_{+}(\tilde{\Omega})$
at $\tilde{\Omega}=\sqrt{2}g$ changes with the rotation detuning
$\Delta$. (d) Demonstration of the ground state and the first three
excited states.}

\label{fig-3peak}
\end{figure}

\textbf{Rabi splitting in the rotating frame} -- We consider the case
that the dipole moment is parallel to the transverse direction, thus
$\xi=0$ and we can choose the polarization direction to satisfy $\vec{\mathfrak{p}}\cdot\hat{\mathrm{e}}_{k\lambda}=|\vec{\mathfrak{p}}|$.
In this case, the excitation number $\hat{N}:=\hat{\sigma}^{z}+\hat{a}_{+}^{\dagger}\hat{a}_{+}+\hat{a}_{-}^{\dagger}\hat{a}_{-}$
is conserved, i.e., {[}$\hat{N},\,\hat{H}]=0$. The ground state is
$|\mathsf{G}\rangle=|\mathsf{g},0,0\rangle$ and the eigen energy
is $E_{\mathsf{G}}=-\Omega/2$. But generally we cannot give an analytical
solution for the whole energy spectrum. When $\omega_{0}=\Omega$,
we can obtain the eigen energy of the first three excited levels {[}Fig.\,\ref{fig-3peak}(d){]},
and they are 
\begin{equation}
E_{0}=\frac{\Omega}{2},\qquad E_{\pm}=\frac{\Omega}{2}\pm\Delta_{g},\label{eq:EigenE}
\end{equation}
where $\Delta_{g}:=\sqrt{\Delta^{2}+2g^{2}}$ (see Appendix \ref{sec:Steady-photon-number}
or Ref.\,\cite{swain_exact_1972}).

Therefore, if we use an external input to drive the ring cavity weakly,
we predict to see Rabi splitting with three resonant peaks \cite{parkins_resonance_1990}.
We consider a probing setup as demonstrated in Fig.\,\ref{fig-setup}.
An external driving laser is input to drive the $k_{+}$-mode of the
ring cavity. The photons in the ring cavity can leak into the output
fiber, and then be probed by photon detectors. 

We use the following master equation to describe the system,
\begin{equation}
\dot{\rho}=i[\rho,\hat{H}+\hat{H}_{\mathrm{d}}(t)]+\sum_{\alpha=+,-}\frac{\gamma}{2}\big(2\hat{a}_{\alpha}\rho\hat{a}_{\alpha}^{\dagger}-\{\rho,\hat{a}_{\alpha}^{\dagger}\hat{a}_{\alpha}\}\big),
\end{equation}
where $\hat{H}_{\mathrm{d}}(t)={\cal E}(e^{i\omega_{\mathrm{d}}t}\hat{a}_{+}+e^{-i\omega_{\mathrm{d}}t}\hat{a}_{+}^{\dagger})$
is the driving term. Here we consider the case $\Omega=\omega_{0}$.
When the driving strength ${\cal E}$ is weak, the cavity modes will
not be excited to states with large photon numbers, and we obtain
(see Appendix \ref{sec:Steady-photon-number}) 
\begin{align}
\overline{n}_{+} & \simeq4{\cal E}^{2}\frac{M}{F},\qquad\overline{n}_{-}\simeq16{\cal E}^{2}\frac{g^{4}}{F},\label{eq:n+-}
\end{align}
where 
\begin{align}
M:= & 4[g^{2}-\tilde{\Omega}(\tilde{\Omega}-\Delta)]^{2}+\gamma^{2}\tilde{\Omega}^{2},\label{eq:n+-ps}\\
F:= & \gamma^{4}\tilde{\Omega}^{2}+8\gamma^{2}[2g^{4}+\tilde{\Omega}^{4}-2g^{2}\tilde{\Omega}^{2}+\Delta^{2}\tilde{\Omega}^{2}]\nonumber \\
 & +16\tilde{\Omega}^{2}[2g^{2}-\tilde{\Omega}^{2}+\Delta^{2}]^{2},\nonumber 
\end{align}
and $\tilde{\Omega}:=\Omega-\omega_{\mathrm{d}}$.

When $\gamma\rightarrow0$, the denominator $F(\tilde{\Omega})$ has
three minima around $\tilde{\Omega}=0$ and $\tilde{\Omega}=\pm\sqrt{\Delta^{2}+2g^{2}}$,
which give rise to three peaks in $\overline{n}_{\pm}(\tilde{\Omega})$
corresponding to the first three excited energy Eq.\,(\ref{eq:EigenE}).
We can also explicitly see that $F(\tilde{\Omega})$ is symmetric
for $\pm\tilde{\Omega}$, i.e., $F(\tilde{\Omega})=F(-\tilde{\Omega})$,
but $M(\tilde{\Omega})$ is not. Therefore, $\overline{n}_{-}(\tilde{\Omega})$
is symmetric for $\pm\tilde{\Omega}$, but $\overline{n}_{+}(\tilde{\Omega})$
is not. We plot the steady photon number $\overline{n}_{\pm}(\tilde{\Omega})$
of the cavity modes in Fig.\,\ref{fig-3peak}. When there is no rotation,
$v_{R}=0$, the steady photon number of the two modes $\overline{n}_{\pm}(\tilde{\Omega})$
are both symmetric with respect to the driving detuning $\pm\tilde{\Omega}$
{[}Fig.\,\ref{fig-3peak}(a){]}. It is worth noticing that when there
is a small rotating velocity, the heights of the side peaks of $\overline{n}_{+}(\tilde{\Omega})$,
which correspond to the mode $\hat{a}_{+}$ being driven, change sensitively
with rotating velocity {[}Fig.\,\ref{fig-3peak}(b){]}.

Since the positions of the side peaks are $\tilde{\Omega}\simeq\pm\sqrt{\Delta^{2}+2g^{2}}$,
we plot the height of $\overline{n}_{+}(\tilde{\Omega}=\sqrt{2}g)$
with respect to the rotation detuning $\Delta=-v_{R}k$ around $\Delta\simeq0$
{[}Fig.\,\ref{fig-3peak}(c){]}. When $\Delta$ is quite small (in
this example, we have $|\Delta/\Omega|=|v_{R}/c|<10^{-5}$), $\overline{n}_{+}(\Delta)$
depend linearly on $\Delta=-v_{R}k$ around $\Delta=0$. From Eqs.\,(\ref{eq:n+-},
\ref{eq:n+-ps}), we obtain the slope of $\overline{n}_{+}(\Delta,\,\tilde{\Omega}=\sqrt{2}g)$
around $\Delta=0$ as
\begin{equation}
\frac{\partial\overline{n}_{+}(\tilde{\Omega},\Delta)}{\partial\Delta}\Big|_{\tilde{\Omega}=\sqrt{2}g}\simeq\frac{64\sqrt{2}g{\cal E}^{2}}{\gamma^{2}(\gamma^{2}+8g^{2})}.
\end{equation}
From these results, we see that the sensitivity of this height change
with respect to the rotating velocity can be controlled by the coupling
strength $g$ and the decay rate $\gamma$. A ring cavity with high
quality promises a sensitive measurement.

In experiments, the steady photon number $\overline{n}_{\pm}$ can
be measured by the average output photon current. In a probe setup
as shown in Fig.\,\ref{fig-setup}, we have $\hat{a}_{\pm,\mathrm{OUT}}=\hat{a}_{\pm,\mathrm{IN}}+\sqrt{\gamma}\hat{a}_{\pm}$
, and the average output photon current is $\langle\hat{a}_{\pm,\mathrm{OUT}}^{\dagger}\hat{a}_{\pm,\mathrm{OUT}}\rangle=\gamma\overline{n}_{\pm}$
\cite{walls_quantum_2008}. This photon current directly characterizes
the steady photon numbers of the cavity modes, and can be measured
by the photon detectors.

\textbf{Summary} -- In conclusion, we have generally studied the quantum
optics for the interaction between light and atom in a non-inertial
reference frame by the approach of quantum field theory in curved
spacetime. For the two-level atom interacting with the quantized EM
field in a rotating ring cavity, our approach built a microscopic
model described by a two-mode JC Hamiltonian. Based on this generalized
JC model, our study predicts that that the heights of the side peaks
in the Rabi splitting show a sensitive linear dependence on the rotating
velocity at low speed. Therefore, this model can not only be utilized
for hybrid optical gyroscope design, but also provide the future development
of quantum gyroscope with a solid physical base to consider the effect
of quantum fluctuation.

This work was supported by the National 973-program (Grants Nos.\,2014CB921403,
2012CB922104 and 2012CB922103), the National Natural Science Foundation
of China (Grants Nos.\,11421063, 11447609, 11404021, 11374095, 11422540,
11434011), and Postdoctoral Science Foundation of China No.\,2013M530516.
S.-W. Li wants to thank Prof. Y. Li and Dr. L. Ge for helpful discussions.

\appendix

\section{spacetime metric \label{sub:Space-time-metric}}

Considering the system rotating along the $z$-axial in the counter-clockwise
direction, we have \cite{menegozzi_theory_1973,chow_ring_1985} 
\begin{align}
t & =T,\nonumber \\
x & =X\cos\Omega t+Y\sin\Omega t,\\
y & =-X\sin\Omega t+Y\cos\Omega t,\nonumber \\
z & =Z.\nonumber 
\end{align}
Here, $X^{\alpha}=(cT,X,Y,Z)$ is the coordinates in the inertial
lab frame, while $x^{\alpha}=(ct,x,y,z)$ is in the co-rotating frame.
We obtain the general coordinate transformation matrix as
\begin{equation}
[\Lambda^{\alpha}{}_{\beta}]=[\frac{\partial x^{\alpha}}{\partial X^{\beta}}]=\left[\begin{array}{cccc}
1\\
\Omega y/c & \cos\Omega t & \sin\Omega t\\
-\Omega x/c & -\sin\Omega t & \cos\Omega t\\
 &  &  & 1
\end{array}\right].
\end{equation}
The spacetime metric $g^{\mu\nu}$ in the co-rotating frame, as a
covariant tensor, can be calculated by 
\begin{align}
[g^{\mu\nu} & ]=[\frac{\partial x^{\mu}}{\partial X^{\alpha}}\cdot\eta^{\alpha\beta}\cdot\frac{\partial x^{\nu}}{\partial X^{\beta}}]\nonumber \\
 & =\left[\begin{array}{cccc}
-1 & -\frac{\Omega y}{c} & \frac{\Omega x}{c}\\
-\frac{\Omega y}{c} & 1-\frac{\Omega^{2}y^{2}}{c^{2}} & \frac{\Omega^{2}xy}{c^{2}}\\
\frac{\Omega x}{c} & \frac{\Omega^{2}xy}{c^{2}} & 1-\frac{\Omega^{2}x^{2}}{c^{2}}\\
 &  &  & 1
\end{array}\right],
\end{align}
 where $\eta^{\mu\nu}=\mathrm{diag}\{-1,1,1,1\}$ is the metric in
the inertial lab frame. And we also have 
\begin{align}
g_{\mu\nu} & =[g^{\mu\nu}]^{-1}\nonumber \\
 & =\left[\begin{array}{cccc}
-1+\frac{\Omega^{2}}{c^{2}}(x^{2}+y^{2}) & -\frac{\Omega y}{c} & \frac{\Omega x}{c}\\
-\frac{\Omega y}{c} & 1\\
\frac{\Omega x}{c} &  & 1\\
 &  &  & 1
\end{array}\right].
\end{align}

\section{Canonical quantization of the rotating EM field \label{sec:Canonical-quantization-of}}

In the canonical quantization of the rotating EM field, it is crucial
to find a proper orthogonal relation of the field operator $\hat{\mathbf{A}}(s,t)$.
This relation can be obtained with the help of a density-flow relation
derived from the equation of motion.

From the equation of $\mathbf{A}(s,t)$
\begin{equation}
[-\partial_{0}^{2}+2\tilde{v}_{R}\partial_{s}\partial_{0}+(1-\tilde{v}_{R}^{2})\partial_{s}^{2}]\mathbf{A}_{i}=0,
\end{equation}
 we obtain the following density-flow relation
\begin{equation}
\partial_{0}[\mathbf{A}^{*}{\cal D}\mathbf{A}]=\partial_{s}[\mathbf{A}^{*}{\cal J}\mathbf{A}],\label{eq:Den-Flow}
\end{equation}
where 
\begin{align*}
\mathbf{A}^{*}{\cal D}\mathbf{A}:= & (\mathbf{A}^{*}\cdot\partial_{0}\mathbf{A}-\mathbf{A}\cdot\partial_{0}\mathbf{A}^{*})\\
 & \qquad-\tilde{v}_{R}(\mathbf{A}^{*}\cdot\partial_{s}\mathbf{A}-\mathbf{A}\cdot\partial_{s}\mathbf{A}^{*}),\\
\mathbf{A}^{*}{\cal J}\mathbf{A}:= & (1-\tilde{v}_{R}^{2})(\mathbf{A}^{*}\cdot\partial_{s}\mathbf{A}-\mathbf{A}\cdot\partial_{s}\mathbf{A}^{*})\\
 & \qquad+\tilde{v}_{R}(\mathbf{A}^{*}\cdot\partial_{0}\mathbf{A}-\mathbf{A}\cdot\partial_{0}\mathbf{A}^{*}).
\end{align*}
Integrating Eq.\,(\ref{eq:Den-Flow}) over the whole space, we obtain
\begin{equation}
\partial_{0}\Big(\int d\mathsf{S}\cdot\int_{0}^{\mathsf{L}}ds[\mathbf{A}^{*}{\cal D}\mathbf{A}]\Big)=0,
\end{equation}
where $\int d\mathsf{S}$ is the integral over the cross section and
gives a constant area $\mathsf{S}$. Then we know that this integral
is a conserved constant independent of time (but still not determined
yet).

Then we can find some orthogonal relations. For the eigen solution
\begin{equation}
\boldsymbol{A}_{k\lambda}(s,t):=\hat{\mathrm{e}}_{k\lambda}\mathsf{a}_{k}(s,t):=\hat{\mathrm{e}}_{k\lambda}\overline{\mathrm{Z}}_{k}e^{iks-i\omega_{k}t},
\end{equation}
 we have the following orthogonal relation 
\begin{align}
\int_{0}^{\mathsf{L}}ds[\boldsymbol{A}_{q\sigma}^{*}{\cal D}\boldsymbol{A}_{k\lambda}] & =-\frac{2i|\overline{\mathrm{Z}}_{k}|^{2}\mathsf{L}}{c}(\omega_{k}+v_{R}k)\,\delta_{kq}\delta_{\lambda\sigma},\nonumber \\
\int_{0}^{\mathsf{L}}ds[\boldsymbol{A}_{q\sigma}{\cal D}\boldsymbol{A}_{k\lambda}] & =\int_{0}^{L}ds[\boldsymbol{A}_{q\sigma}^{*}{\cal D}\boldsymbol{A}_{k\lambda}^{*}]=0.\label{eq:orthog}
\end{align}
Notice that no matter $k>0$ and $k<0$, we both have $\omega_{k}+v_{R}k=c|k|$.

With the help of the above orthogonal relation, for the field operator
\[
\hat{\mathbf{A}}(s,t)=\sum_{k,\lambda}\hat{\mathrm{e}}_{k\lambda}\overline{\mathrm{Z}}_{k}[\hat{a}_{k\lambda}e^{iks-i\omega_{k}t}+\hat{a}_{k\lambda}^{\dagger}e^{-iks+i\omega_{k}t}],
\]
 we have
\begin{align}
\int_{0}^{L}ds[\boldsymbol{A}_{k\lambda}^{*}{\cal D}\hat{\mathbf{A}}] & =-2i|\overline{\mathrm{Z}}_{k}|^{2}\mathsf{L}|k|\cdot\hat{a}_{k\lambda},\nonumber \\
\int_{0}^{L}ds[\boldsymbol{A}_{k\lambda}{\cal D}\hat{\mathbf{A}}] & =2i|\overline{\mathrm{Z}}_{k}|^{2}\mathsf{L}|k|\cdot\hat{a}_{k\lambda}^{\dagger}.\label{eq:a_k}
\end{align}

Notice that indeed the above terms in the integrals can be written
as
\begin{align*}
\boldsymbol{A}_{k\lambda}^{*}{\cal D}\hat{\mathbf{A}} & =\mathbf{a}_{k\lambda}^{*}\cdot(\partial_{0}-\tilde{v}_{R}\partial_{s})\hat{\mathbf{A}}-\hat{\mathbf{A}}\cdot(\partial_{0}-\tilde{v}_{R}\partial_{s})\mathbf{a}_{k\lambda}^{*}\\
 & =\frac{\mathsf{a}_{k}^{*}{\cal \hat{E}}^{\lambda}}{c\varepsilon_{0}}-i|k|\mathsf{a}_{k}^{*}\hat{\mathbf{A}}_{\lambda},\\
\boldsymbol{A}_{k\lambda}{\cal D}\hat{\mathbf{A}} & =\mathbf{a}_{k\lambda}\cdot(\partial_{0}-\tilde{v}_{R}\partial_{s})\hat{\mathbf{A}}-\hat{\mathbf{A}}\cdot(\partial_{0}-\tilde{v}_{R}\partial_{s})\mathbf{a}_{k\lambda}\\
 & =\frac{\mathsf{a}_{k}{\cal \hat{E}}^{\lambda}}{c\varepsilon_{0}}+i|k|\mathsf{a}_{k}\hat{\mathbf{A}}_{\lambda}.
\end{align*}
Then, together with the canonical quantization conditions $[\hat{\mathbf{A}}_{\lambda}(s,t),\,\hat{{\cal E}}^{\sigma}(s',t)]=i\hbar\delta_{\lambda}^{\sigma}\delta(s-s')\cdot\mathsf{A}^{-1}$
and $[\hat{\mathbf{A}}_{\lambda}(s,t),\,\hat{\mathbf{A}}_{\sigma}(s',t)]=[\hat{{\cal E}}^{\lambda}(s,t),\,\hat{{\cal E}}^{\sigma}(s',t)]=0$,
we can calculate $[\hat{a}_{k\lambda},\hat{a}_{q\sigma}^{\dagger}]$
from Eq.\,(\ref{eq:a_k}) as follows
\begin{align*}
 & [\,2|\overline{\mathrm{Z}}_{k}|^{2}\mathsf{L}|k|\cdot\hat{a}_{k\lambda},\,2|\overline{\mathrm{Z}}_{q}|^{2}\mathsf{L}|q|\cdot\hat{a}_{q\sigma}^{\dagger}]\\
= & \iint dsds'\big[\boldsymbol{A}_{k\lambda}^{*}(s,t){\cal D}\hat{\mathbf{A}}(s,t),\,\boldsymbol{A}_{q\sigma}(s',t){\cal D}\hat{\mathbf{A}}(s',t)\big]\\
= & \iint dsds'\big[\frac{\mathsf{a}_{k}^{*}\hat{{\cal E}}^{\lambda}(s)}{c\varepsilon_{0}}-i|k|\mathsf{a}_{k}^{*}\hat{\mathbf{A}}_{\lambda}(s),\,\frac{\mathsf{a}_{q}\hat{{\cal E}}^{\sigma}(s')}{c\varepsilon_{0}}+i|q|\mathsf{a}_{q}\hat{\mathbf{A}}_{\sigma}(s')\big]\\
= & \frac{2\hbar\mathsf{L}|k|}{c\varepsilon_{0}\mathsf{S}}\cdot|\overline{\mathrm{Z}}_{k}|^{2}\cdot\delta_{\lambda}^{\sigma}\delta_{kq}.
\end{align*}
Therefore, we choose the normalization constant $\overline{\mathrm{Z}}_{k}$
to be
\begin{equation}
\overline{\mathrm{Z}}_{k}=[\frac{\hbar}{2\varepsilon_{0}V(\omega_{k}+v_{R}k)}]^{\frac{1}{2}}=[\frac{\hbar}{2\varepsilon_{0}Vc|k|}]^{\frac{1}{2}},
\end{equation}
where $V:=\mathsf{L}\cdot\mathsf{S}$ is the volume of the quasi-1D
ring, so that we obtain the bosonic commutation relation $[\hat{a}_{k\lambda},\hat{a}_{q\sigma}^{\dagger}]=\delta_{\lambda\sigma}\delta_{kq}$.
With the same method, we can also check $[\hat{a}_{k\lambda},\,\hat{a}_{q\sigma}]=[\hat{a}_{k\lambda}^{\dagger},\,\hat{a}_{q\sigma}^{\dagger}]=0$.

After we have got the normalization constant $\overline{\mathrm{Z}}_{k}$,
it is straightforward to verify that the Hamiltonian of the rotating
EM field is
\begin{align*}
\hat{H}_{\mathrm{EM}} & =\int dV\,{\cal H}\\
 & =\int dV\,\big[\frac{1}{2}\varepsilon_{0}(\partial_{t}\hat{\mathbf{A}})^{2}+\frac{1-\tilde{v}_{R}^{2}}{2\mu_{0}}(\partial_{s}\hat{\mathbf{A}})^{2}\big]\\
 & =\sum_{k,\lambda}\frac{1}{2}\hbar\omega_{k}(\hat{a}_{k\lambda}^{\dagger}\hat{a}_{k\lambda}+\hat{a}_{k\lambda}\hat{a}_{k\lambda}^{\dagger}).
\end{align*}

\section{Hamiltonian description of a charged particle in the co-rotating
EM field \label{sec:Hamiltonian-description}}

The Hamiltonian equation can be derived from the following variation,
\begin{equation}
\delta\int Ld\tau=\delta\int\tilde{L}dt=\delta\int(P_{i}\frac{dx^{i}}{dt}-H)dt=0,
\end{equation}
where $H[x_{i}(t),\,P^{i}(t)]$ is the functional of independent variables
$x^{i}(t)$ and $P_{i}(t)$. The above variation gives
\begin{align}
0= & \delta\int\Big(P_{i}\frac{dx^{i}}{dt}-H[x^{i},\,P_{i}]\Big)dt\\
= & P_{i}\delta x^{i}\Big|_{A}^{B}+\delta\int(\frac{dx^{i}}{dt}-\frac{\partial H}{\partial P_{i}})\delta P_{i}+(\frac{dP_{i}}{dt}+\frac{\partial H}{\partial x^{i}})\delta x^{i}.\nonumber 
\end{align}
$A$ and $B$ are the initial and final points of the trajectories
of $x^{i}(t)$ which are fixed, and thus the first term is zero. Since
$\delta x^{i}$ and $\delta P_{i}$ are independent variables, the
above variation leads to the Hamilton equations
\begin{equation}
\frac{dx^{i}}{dt}=\frac{\partial H}{\partial P_{i}},\quad\frac{dP_{i}}{dt}=-\frac{\partial H}{\partial x^{i}}.
\end{equation}
 Here we have utilized the following relation
\begin{align}
\delta\int P_{i}v^{i}dt & =\int\Big[v^{i}\delta P_{i}+P_{i}\delta(\frac{dx^{i}}{dt})\Big]dt\nonumber \\
 & =\int\Big[v^{i}\delta P_{i}+P_{i}\frac{d}{dt}(\delta x^{i})\Big]dt\\
 & =P_{i}\delta x^{i}\Big|_{A}^{B}+\int\Big[\frac{d}{dt}x^{i}\cdot\delta P_{i}-\frac{d}{dt}P_{i}\cdot\delta x^{i}\Big]dt.\nonumber 
\end{align}

Therefore, for a charged particle in the co-rotating EM field, we
have
\begin{align}
P_{i} & =\frac{\partial\tilde{L}}{\partial v^{i}}=\Gamma mv_{i}+eA_{i}:=p_{i}+eA_{i},\nonumber \\
H & =P_{i}v^{i}-\tilde{L}=\Gamma m(v_{i}v^{i}+\frac{c^{2}}{\Gamma^{2}})-eA_{0}v^{0}\label{eq:Ham-e}\\
 & =-\Gamma mv_{0}v^{0}-eA_{0}v^{0}.\nonumber 
\end{align}
Here we denote $v^{\mu}:=dx^{\mu}/dt$, $v_{\mu}:=dx_{\mu}/dt$. And
we denote $p_{i}:=\Gamma mv_{i}$ as the mechanical momentum. Here
we take $x^{\mu}=(ct,x_{1},x_{2},x_{3})$ and $x_{\mu}=g_{\mu\nu}x^{\nu}$,
thus we have $v^{0}=c$. 

Further, we still need to replace $\Gamma mv_{0}$ by the canonical
momentum $P_{i}$ in the Hamiltonian $H$. We emphasize that here
we only have got the relation between $P_{i},\,p_{i}$ and $v_{i}$
from Eq.\,(\ref{eq:Ham-e}), but we did not have the definition of
 $P_{0},\,p_{0}$, so we cannot just naively replace $\Gamma mv_{0}$
as $p_{0}$. To find the relation between $\Gamma mv_{0}$ and $P_{i},\,p_{i}$,
we should notice the following two relations,

\begin{align}
\Gamma^{2}m^{2}v_{i}v^{i} & =\Gamma^{2}m^{2}v_{i}(g^{i0}v_{0}+g^{ij}v_{j})\\
 & =g^{i0}p_{i}\cdot\Gamma mv_{0}+g^{ij}p_{i}p_{j}\nonumber \\
\Gamma^{2}m^{2}v_{i}v^{i} & =m^{2}c^{2}\cdot\frac{v_{i}v^{i}}{-v_{0}v^{0}-v_{i}v^{i}}\\
 & =-m^{2}c^{2}-\Gamma mv_{0}\cdot\Gamma mv^{0}\nonumber \\
 & =-g^{00}(\Gamma mv_{0})^{2}-g^{0i}p_{i}\cdot(\Gamma mv_{0})-m^{2}c^{2}\nonumber 
\end{align}
Therefore, we obtain an equation about $\Gamma mv_{0}$, i.e.,
\begin{equation}
g^{00}(\Gamma mv_{0})^{2}+2g^{0i}p_{i}\cdot(\Gamma mv_{0})+g^{ij}p_{i}p_{j}+m^{2}c^{2}=0,
\end{equation}
which leads to the solution ($\Gamma mv_{0}$ should be negative,
so the other positive solution is invalid) 
\[
\Gamma mv_{0}=\frac{g^{0i}p_{i}-\sqrt{(g^{0i}p_{i})^{2}-g^{00}(g^{ij}p_{i}p_{j}+m^{2}c^{2})}}{-g^{00}}.
\]
Therefore, we have
\[
H_{\mathrm{e}}=\frac{v^{0}}{g^{00}}[g^{0i}p_{i}-\sqrt{(g^{0i}p_{i})^{2}-g^{00}(g^{ij}p_{i}p_{j}+m^{2}c^{2})}]-ev^{0}A_{0},
\]
where $p_{i}=P_{i}-eA_{i}$.

\section{Steady photon number \label{sec:Steady-photon-number}}

For the JC-model for a two-level atom in a rotating ring cavity 
\begin{align}
\hat{H}= & \frac{\hbar\Omega}{2}\hat{\sigma}^{z}+\xi\hat{\sigma}^{y}+\hbar\omega_{+}\hat{a}_{+}^{\dagger}\hat{a}_{+}+\hbar\omega_{-}\hat{a}_{-}^{\dagger}\hat{a}_{-}\\
 & \qquad+g\hat{\sigma}^{+}(\hat{a}_{+}+\hat{a}_{-})+g^{*}\hat{\sigma}^{-}(\hat{a}_{+}^{\dagger}+\hat{a}_{-}^{\dagger}),\nonumber 
\end{align}
we can directly verify that the ground state is $|\mathsf{g},0,0\rangle$
and the eigen energy is $E_{\mathsf{G}}=-\Omega/2$. 

The excitation number $\hat{N}:=\hat{\sigma}^{z}+\hat{a}_{+}^{\dagger}\hat{a}_{+}+\hat{a}_{-}^{\dagger}\hat{a}_{-}$
is always conserved, i.e., {[}$\hat{N},\,\hat{H}]=0$. Thus, the single
excitation subspace, which is spanned by $|\mathsf{e},0,0\rangle,\,|\mathsf{g},1,0\rangle,\,|\mathsf{g},0,1\rangle$,
is closed, i.e.,
\begin{align}
\hat{H}|\mathsf{e},0,0\rangle & =\frac{\Omega}{2}|\mathsf{e},0,0\rangle+g|\mathsf{g},1,0\rangle+g|\mathsf{g},0,1\rangle\nonumber \\
\hat{H}|\mathsf{g},1,0\rangle & =g|\mathsf{e},0,0\rangle+(\frac{\omega_{0}}{2}+\Delta)|\mathsf{g},1,0\rangle\\
\hat{H}|\mathsf{g},0,1\rangle & =g|\mathsf{e},0,0\rangle+(\frac{\omega_{0}}{2}-\Delta)|\mathsf{g},0,1\rangle\nonumber 
\end{align}

When $\Omega=\omega_{0}$, the above eigen equations can be diagonalized
exactly and we can obtain the eigen energy states in the single excitation
subspace. The eigen energies are $E_{0}=\Omega/2$ and $E_{\pm}=\Omega/2\pm\Delta_{g}$,
where $\Delta_{g}:=\sqrt{\Delta^{2}+2g^{2}}$, and the eigenstates
are
\begin{alignat}{1}
|E_{0}\rangle= & \frac{1}{Z_{0}}\Big[\Delta|\mathsf{e},0,0\rangle-g|\mathsf{g},1,0\rangle+g|\mathsf{g},0,1\rangle\Big],\\
|E_{\pm}\rangle= & \frac{1}{Z_{\pm}}\Big[g(\Delta\pm\Delta_{g})|\mathsf{e},0,0\rangle+\frac{1}{2}(\Delta\pm\Delta_{g})^{2}|\mathsf{g},1,0\rangle\nonumber \\
 & +|g|^{2}|\mathsf{g},0,1\rangle\Big],\nonumber 
\end{alignat}
where $Z_{0}$ and $Z_{\pm}$ are normalization constants.

When we use an external laser to drive the $\hat{a}_{+}$ mode, the
master equation of the system is
\begin{equation}
\dot{\rho}=i[\rho,\hat{H}+\hat{H}_{\mathrm{d}}(t)]+\sum_{\alpha=+,-}\frac{\gamma}{2}\big(2\hat{a}_{\alpha}\rho\hat{a}_{\alpha}^{\dagger}-\{\rho,\hat{a}_{\alpha}^{\dagger}\hat{a}_{\alpha}\}\big),
\end{equation}
where $\hat{H}_{\mathrm{d}}={\cal E}(e^{i\omega_{d}t}\hat{a}_{+}+e^{-i\omega_{d}t}\hat{a}_{+}^{\dagger})$
is the driving term. We make a unitary transformation by $\exp[i\omega_{\mathrm{d}}(\hat{a}_{+}^{\dagger}\hat{a}_{+}+\hat{a}_{-}^{\dagger}\hat{a}_{-}+\hat{\sigma}^{z}/2)]$,
and the equation becomes time independent. Then we obtain equations
for observable expectations as follows 
\begin{align*}
0 & =-i(\tilde{\omega}_{+}-i\frac{\gamma}{2})\alpha_{+}-ig\langle\hat{\sigma}^{-}\rangle-i{\cal E}\\
0 & =-i(\tilde{\omega}_{-}-i\frac{\gamma}{2})\alpha_{-}-ig\langle\hat{\sigma}^{-}\rangle\\
0 & =-i(\tilde{\omega}_{+}-i\frac{\gamma}{2})S_{+}-i{\cal E}\langle\hat{\sigma}^{z}\rangle+ig\langle\hat{\sigma}^{-}\rangle\\
0 & =-i(\tilde{\omega}_{-}-i\frac{\gamma}{2})S_{-}+ig\langle\hat{\sigma}^{-}\rangle\\
0 & =-i(\tilde{\omega}_{+}-\tilde{\Omega}-i\frac{\gamma}{2})Z_{+}-i{\cal E}\langle\hat{\sigma}^{+}\rangle-i\frac{g}{2}(\langle\hat{\sigma}^{z}\rangle+1)\\
0 & =-i(\tilde{\omega}_{-}-\tilde{\Omega}-i\frac{\gamma}{2})Z_{-}-i\frac{g}{2}(\langle\hat{\sigma}^{z}\rangle+1)\\
0 & =-i\tilde{\Omega}\langle\hat{\sigma}^{-}\rangle+ig(S_{+}+S_{-})\\
0 & =(Z_{+}-Z_{+}^{*})+(Z_{-}-Z_{-}^{*})
\end{align*}
Here we denote $Z_{\pm}:=\langle\hat{\sigma}^{+}\hat{a}_{\pm}\rangle,\,S_{\pm}:=\langle\hat{\sigma}^{z}\hat{a}_{\pm}\rangle,\,\alpha_{\pm}:=\langle\hat{a}_{\pm}\rangle$
and $\tilde{\omega}_{\pm}:=\omega_{\pm}-\omega_{\mathrm{d}}$, $\tilde{\Omega}:=\Omega-\omega_{\mathrm{d}}$.
In this equation we have omitted terms of higher orders, like $\langle\hat{\sigma}^{z}\hat{a}_{\pm}^{\dagger}\hat{a}_{\pm}\rangle,\,\langle\hat{\sigma}^{z}\hat{a}_{+}^{\dagger}\hat{a}_{-}\rangle$,
so the above linear equations become complete. This approximation
requires that the driving strength is weak thus the excitation is
very low.

From this set of linear equations, we obtain the solution of $\alpha_{\pm}=\langle\hat{a}_{\pm}\rangle$,
and then we obtain $\overline{n}_{\pm}\simeq|\alpha_{\pm}|^{2}$ as
follows
\begin{align}
\overline{n}_{+}\simeq & |\alpha_{+}|^{2}=\frac{4{\cal E}^{2}[MF^{2}D+o({\cal E}^{2})]}{(FD+4{\cal E}^{2}G)^{2}},\label{eq:N+-}\\
\overline{n}_{-}\simeq & |\alpha_{-}|^{2}=\frac{16{\cal E}^{2}g^{4}\cdot F^{2}D}{(FD+4{\cal E}^{2}G)^{2}},\nonumber 
\end{align}
where 
\begin{align*}
M:= & 4(g^{2}-\tilde{\Omega}\tilde{\omega}_{-})^{2}+\gamma^{2}\tilde{\Omega}^{2},\\
D:= & \gamma^{2}+4\Delta^{2},\\
F:= & \gamma^{4}\tilde{\Omega}^{2}+4\gamma^{2}[2g^{4}+(\tilde{\omega}_{+}\tilde{\Omega}-g^{2})^{2}+(\tilde{\omega}_{-}\tilde{\Omega}-g^{2})^{2}]\\
 & +16\tilde{\Omega}^{2}[2g^{2}-\tilde{\Omega}^{2}+\Delta^{2}]^{2},\\
G:= & 16\Delta^{2}\big[(2g^{2}-\tilde{\Omega}^{2})\tilde{\omega}_{-}^{2}+2g^{2}\tilde{\Omega}\Delta\big]+\gamma^{4}(2g^{2}-\tilde{\Omega}^{2})\\
 & +4\gamma^{2}\big[(2g^{2}-\tilde{\Omega}^{2})\Delta^{2}+(2g^{2}-\tilde{\Omega}^{2})\tilde{\omega}_{-}^{2}+2g^{2}\tilde{\Omega}\Delta\big].
\end{align*}
When the driving strength ${\cal E}$ is weak, we omit terms of $o({\cal E}^{2})$
in Eq.\,(\ref{eq:N+-}) and obtain
\begin{align}
\overline{n}_{+} & \simeq\frac{4{\cal E}^{2}M}{F},\qquad\overline{n}_{-}\simeq\frac{16{\cal E}^{2}g^{4}}{F}.
\end{align}
For the case that the atom is resonant with the resonant frequency,
i.e., $\Omega=\omega_{0}$, we have $\tilde{\omega}_{\pm}=\tilde{\Omega}\pm\Delta$,
and we obtain
\begin{align*}
M:= & 4[g^{2}-\tilde{\Omega}(\tilde{\Omega}-\Delta)]^{2}+\gamma^{2}\tilde{\Omega}^{2},\\
F:= & \gamma^{4}\tilde{\Omega}^{2}+8\gamma^{2}[2g^{4}+\tilde{\Omega}^{4}-2g^{2}\tilde{\Omega}^{2}+\Delta^{2}\tilde{\Omega}^{2}]\\
 & +16\tilde{\Omega}^{2}[2g^{2}-\tilde{\Omega}^{2}+\Delta^{2}]^{2}
\end{align*}
This is what we have shown in the text.

The above master equation can be also solved numerically by setting
a cutoff on the Hilbert dimension of the cavity modes. We find that
when the driving strength ${\cal E}$ is weak, the above analytical
results of the steady photon number show well accordance with the
numerical calculations.
\end{document}